\documentclass[aps,pra,twocolumn,showpacs,preprintnumbers,amsmath,amssymb,superscriptaddress,10pt]{revtex4-1}

\usepackage{amsmath,amssymb}
\usepackage{bm}
\usepackage[latin1]{inputenc} 
\usepackage{dcolumn}
\usepackage{graphicx}
\usepackage{SIunits}
\usepackage{ae}
\usepackage{subfig}
\usepackage{color} 
\usepackage{tikz}

\newcommand{\matrise}[1]{\begin{bmatrix} #1 \end{bmatrix}}
\newcommand{\diff}[0]{\text{d}}

\newcommand{\re}{\mathrm{Re}\,}
\newcommand{\im}{\mathrm{Im}\,}

\newcommand{\vek}[1]{\boldsymbol{\mathbf{#1}}}
\newcommand{\vekh}[1]{\hat{{\boldsymbol{\mathbf{#1}}}}}

\newcommand{\be}{\begin{equation}}
\newcommand{\ee}{\end{equation}}
\newcommand{\ba}{\begin{align}}
\newcommand{\ea}{\end{align}}
\newcommand{\e}[1]{\text{e}^{#1}}

\begin{document}

\title{Diamagnetism and the dispersion of the magnetic permeability}

\author{Christopher A. Dirdal}
\affiliation{Department of Electronic Systems, NTNU -- Norwegian University of Science and Technology, NO-7491 Trondheim, Norway}
\author{Johannes Skaar}
\affiliation{Department of Electronic Systems, NTNU -- Norwegian University of Science and Technology, NO-7491 Trondheim, Norway}
\affiliation{Department of Technology Systems, University of Oslo, Box 70, NO-2027 Kjeller, Norway}
\email{johannes.skaar@its.uio.no}

\date{\today}

\begin{abstract}
It is well known that the usual Kramers--Kronig relations for the relative permeability function $\mu(\omega)$ are not compatible with diamagnetism ($\mu(0)<1$) and a positive imaginary part ($\im\mu(\omega)>0$ for $\omega>0$). We demonstrate that a certain physical meaning can be attributed to $\mu$ for all frequencies, and that in the presence of spatial dispersion, $\mu$ does not necessarily tend to 1 for high frequencies $\omega$ and fixed wavenumber $\vek k$. Taking the asymptotic behavior into account, diamagnetism can be compatible with Kramers--Kronig relations even if the imaginary part of the permeability is positive. We provide several examples of diamagnetic media and metamaterials for which $\mu(\omega,\vek k)\not\to 1$ as $\omega\to\infty$.
\end{abstract}

\maketitle

\section{Introduction}
With recent advances in metamaterial research, there is a renewed interest in the properties of the magnetic permeability function. The permeability function extracted by homogenization methods may show peculiar properties, such as a negative imaginary part, anomalous dispersion effects, and diamagnetism \cite{koschny03,Silveirinha11}. This has lead to investigations on the properties of the magnetic permeability \cite{koschny03,Silveirinha11,efros04,silveirinha09,alu11,pitaevskii12,Yaghjian2013374}.

A related problem is the well known paradox that diamagnetism is not compatible with a permeability function that satisfies the Kramers--Kronig relations and has a positive imaginary part. Letting $\mu(\omega)$ be the relative permeability, and considering zero frequency in the Kramers--Kronig relation, one has
\be\label{KK0}
\mu(0)-1 = \frac{2}{\pi}\int_0^\infty\frac{\im\mu(\omega)\diff\omega}{\omega}.
\ee
Apparently, \eqref{KK0} predicts that $\mu(0)<1$ is not possible for media with $\im\mu(\omega)>0$. It has been argued that the Kramers--Kronig relations should be modified \cite{Silveirinha11,landau_lifshitz_edcm} to avoid this problem. It has also been argued that the magnetic permeability can have a negative imaginary part, at least for high frequencies \cite{Silveirinha11,pitaevskii12,Yaghjian2013374,martin67,markel08}.

In this article we show that Maxwell's equations and the fundamental principle of causality do not require the magnetic relative permeability to approach unity for high frequencies and fixed wavenumbers (Sec. \ref{sec:elmagpar}). Causality is not violated, as the requirement $\mu\to 1$ for high frequencies is only necessary under eigenmodal propagation (in the absence of sources in the medium), where $k$ and $\omega$ are connected by the dispersion relation $k=\sqrt{\epsilon\mu}\omega/c$. Here $\epsilon$ is the relative permittivity, $c=1/\sqrt{\epsilon_0\mu_0}$ is the vacuum light velocity, and $\epsilon_0$ and $\mu_0$ are the vacuum permittivity and permeability, respectively. We consider the ambiguity in associating induced currents with the electric polarization or magnetization, and describe, somewhat artificially, a simple homogeneous conductor or superconductor as concrete examples of media possessing diamagnetism (Sec. \ref{sec:homcond}). We find that their relative permeabilities do not tend to unity for high frequencies and fixed wavenumbers. Finally, we evaluate the relative permeability of 1d and 2d metamaterial examples with conducting inclusions, and demonstrate analytically and numerically that it does not tend to unity for high frequencies (Sec. \ref{sec:met}).

Our findings do not contradict previous results \cite{Silveirinha11,pitaevskii12,landau_lifshitz_edcm,martin67}, but rather imply that diamagnetism is possible even if $\im\mu>0$ for all positive frequencies. Here we note that the definition of $\mu$ can be extrapolated to all frequencies, such that the integrals in the Kramers--Kronig relations do not have to be truncated. The function $\mu$ will certainly not have its usual interpretation as a local parameter for all frequencies, but rather be a quantity relating the averaged magnetic moment of the material's unit cell to the averaged magnetic field.

For ease of notation we will denote the relative permittivity and permeability by $\epsilon$ and $\mu$, respectively, while the absolute permittivity and permeability will be expressed $\epsilon\epsilon_0$ and $\mu\mu_0$.

\section{Electromagnetic parameters}\label{sec:elmagpar}
Following the classic treatments \cite{landau_lifshitz_edcm,agranovich84}, we consider a passive and time-shift invariant medium, and formulate electromagnetism in frequency--wavenumber space. The Amp\-ere--Maxwell's law and Faraday's law can be written 
\begin{subequations}\label{maxwell}
\begin{align}
 \frac{1}{\mu_0}i\vek k\times\vek B & =-i\omega\epsilon_0\vek E+\vek J+\vek J_\text{ext}, \\
 i\vek k\times\vek E & =i\omega\vek B. \label{faraday}
\end{align}
\end{subequations}
Here $\vek J_\text{ext}$ is the external source, and $\vek J$ is the induced current density. We consider a single spatial Fourier component of the source, i.e., $\vek J_\text{ext}=\bar{\vek J}_\text{ext}\exp(i\vek k\cdot\vek r)$ with constant $\bar{\vek J}_\text{ext}$. In a homogeneous medium, $\vek E$ and $\vek B$ are the electric field and magnetic flux density, respectively. In a periodic metamaterial, the Maxwell equations \eqref{maxwell} result from an averaging of the corresponding microscopic equations \cite{silveirinha09,alu11,Yaghjian2013374,dirdal2017}. The macroscopic electric and magnetic fields $\vek E$ and $\vek B$ are then defined by the so-called fundamental Floquet modes of the microscopic fields $\vek e$ and $\vek b$, respectively:
\begin{subequations}\label{averagefield1}
\begin{align}
    \vek E(\vek r) &= \frac{\text{e}^{i\mathbf{k \cdot r}}}{V}\int_V \vek e(\mathbf{r'}) \text{e}^{-i\mathbf{k \cdot r'}} \diff^3 r', \\
   \vek B(\vek r) &= \frac{\text{e}^{i\mathbf{k \cdot r}}}{V}\int_V \vek b(\mathbf{r'}) \text{e}^{-i\mathbf{k \cdot r'}} \diff^3 r',
\end{align}
\end{subequations}
where integration is over a unit cell volume $V$.
Similarly, the macroscopic induced current density is
\begin{equation}
   \vek J(\vek r) = \frac{\text{e}^{i\mathbf{k \cdot r}}}{V}\int_V \vek j(\mathbf{r'}) \text{e}^{-i\mathbf{k \cdot r'}} \diff^3 r'.
\end{equation}

The induced current can be decomposed in several ways. One option is to decompose it in terms of polarization and magnetization terms,
\be\label{indJPM}
\vek J = -i\omega\vek P + i\vek k\times\vek M.
\ee
In \eqref{indJPM} the two terms do not have to be defined from a multipole expansion; thus there is no loss of generality. Of course, the term $i\vek k\times\vek M$ can only contain transverse parts of the induced current, so the longitudinal part must be contained in $-i\omega\vek P$.

Assuming a linear medium, permittivity and permeability dyadics $\vek\epsilon$ and $\vek\mu$ are defined from
\begin{subequations}\label{defPM}
\begin{align}
\vek P &= \epsilon_0(\vek\epsilon-1)\vek E, \\
\mu_0\vek M &= (1-\vek\mu^{-1})\vek B, \label{defMB}
\end{align}
\end{subequations}
respectively. In general, both $\vek\epsilon$ and $\vek\mu$ are dependent on $\omega$ and $\vek k$, i.e., they are temporally and spatially dispersive.

By introducing auxiliary fields $\vek D=\epsilon_0\vek E+\vek P=\vek\epsilon\epsilon_0\vek E$ and $\vek H=\vek B/\mu_0-\vek M$, we obtain from \eqref{maxwell} the Maxwell equations
\begin{subequations}\label{maxwellaux}
\begin{align}
 i\vek k\times\vek H & =-i\omega\vek\epsilon\epsilon_0\vek E+\vek J_\text{ext}, \\
 i\vek k\times\vek E & =i\omega\vek\mu\mu_0\vek H. \label{faradayaux}
\end{align}
\end{subequations}
For simplicity, assume $\vek\mu$ is a scalar $\mu$, and $\vek\epsilon$ in the form \cite{landau_lifshitz_edcm}
\be\label{epstensor}
\vek\epsilon = \matrise{\epsilon_\parallel & 0 & 0 \\ 0 & \epsilon_\perp & 0 \\ 0 & 0 & \epsilon_\perp},
\ee
expressed in an orthogonal basis where the first unit vector is $\vek k/k$. By combining the two Maxwell equations \eqref{maxwellaux}, we obtain the solutions
\begin{subequations}\label{maxwellsol}
\begin{align}
\vek B &=\frac{i\mu\mu_0\vek k\times\vek J_{\text{ext}\perp}}{k^2-\frac{\omega^2}{c^2}\epsilon_\perp\mu}, \label{BperpJ}\\
\vek E_\perp &=\frac{i\omega\mu\mu_0\vek J_{\text{ext}\perp}}{k^2-\frac{\omega^2}{c^2}\epsilon_\perp\mu}, \label{EperpJ}\\
\vek E_\parallel &= \frac{\vek J_{\text{ext}\parallel}}{i\omega\epsilon_\parallel\epsilon_0}. \label{EparJ}
\end{align}
\end{subequations}
In \eqref{maxwellsol} the source $\vek J_\text{ext}$ and field $\vek E$ are decomposed into their components parallel $\parallel$ and perpendicular $\perp$ to $\vek k$. 

As $\omega\to\infty$ the fields do not feel the presence of the medium, so the solutions \eqref{maxwellsol} must be the same as if we set $\epsilon$ and $\mu$ equal to unity in the expressions. Considering \eqref{EparJ} this immediately gives that $\epsilon_\parallel\to 1$. Also, for fixed $k$, \eqref{EperpJ} means $\epsilon_\perp\to 1$ (excluding the possibility $\epsilon_\perp\mu\to 0$). Remarkably, we do not get any condition for the asymptotic behavior of $\mu$. Indeed, for fixed $k$, the expressions become independent of $\mu$ in the limit $\omega\to\infty$. Thus the relative permeability is not required to approach unity for high frequencies.

Nevertheless, even though $\mu$ does not necessarily approach unity, the magnetization $\vek M$ vanishes in the limit $\omega\to\infty$. This is a result of the fact that according to \eqref{BperpJ}, the magnetic field vanishes in this limit.

For eigenmodal propagation, $k^2=\epsilon_\perp\mu\omega^2/c^2$, the situation is different. To see this, let a current source plane be located somewhere in the infinite medium. In the limit $\omega\to\infty$ the generated waves must have the same phase velocity as if the medium were not present. In other words, $\epsilon_\perp\mu\to 1$ under eigenmodal propagation. This ensures relativistic causality, i.e., the front of a wave packet propagates at the speed of light in vacuum.

From \eqref{indJPM}, any transformation $\vek P\to\vek P'$ and $\vek M\to\vek M'$ satisfying
\be\label{magnpol}
-i\omega\vek P' + i\vek k\times\vek M' = -i\omega\vek P + i\vek k\times\vek M
\ee
will leave the induced current density $\vek J$, and therefore also the fundamental fields $\vek E$ and $\vek B$, unchanged. This means that the relative permittivity and permeability are not uniquely defined by \eqref{indJPM}-\eqref{defPM}. We can define primed relative permittivity $\epsilon'$ and permeability $\mu'$ from $\vek P'$ and $\vek M'$, similarly to \eqref{defPM}. Substituting the unprimed and primed version of \eqref{defPM} into \eqref{magnpol}, and eliminating $\vek B$ with \eqref{faraday},
\begin{align}
  (\omega^2/c^2)(\epsilon' &-1)\vek E - \vek k\times (1-\mu'^{-1})\vek k\times\vek E \nonumber\\
= (\omega^2/c^2)(\epsilon &-1)\vek E - \vek k\times (1-\mu^{-1})\vek k\times\vek E. \label{transfPM}
\end{align}
Imagine now that a certain set $\epsilon$ and $\mu$ of a medium is known. Then \eqref{transfPM} predicts the existence of another set of parameters $\epsilon'$ and $\mu'$, which is equivalent to the first set. By considering the components of \eqref{transfPM} parallel and perpendicular to $\vek k$, we find
\begin{subequations}\label{transf}
\begin{align}
 \epsilon_\parallel'&= \epsilon_\parallel, \label{epsl}\\ 
 \epsilon'_\perp + \frac{k^2c^2}{\omega^2}(1-\mu'^{-1}) &= \epsilon_\perp + \frac{k^2c^2}{\omega^2}(1-\mu^{-1}), \label{epst} 
\end{align}
\end{subequations}
respectively. The relation \eqref{transf} is well known in literature, although it is usually specialized to the case $\mu'=1$. This case, which is called the Landau--Lifshitz formulation, is particularly useful in the presence of spatial dispersion, where there is no set of local (independent of $\vek k$) parameters $\epsilon$ and $\mu$ \cite{landau_lifshitz_edcm,agranovich84}. Then it is convenient to specify the medium properties by a single (but nonlocal) quantity $\epsilon'$. For spatially nondispersive media where $\epsilon$ and $\mu$ are independent of $\vek k$, it is often more convenient to retain these two parameters, as they are much simpler to use in practical situations formulated in the spatial domain. Since there is not always a set $\epsilon$ and $\mu$ such that the parameters are independent of $\vek k$, we allow all parameters in \eqref{transf}, $\epsilon$, $\mu$, $\epsilon'$, and $\mu'$, to depend on $(\omega,\vek k)$ although not explicitly specified.

Apparently, a medium described by a given set of parameters $\epsilon$ and $\mu$ is equally well described by any other set of parameters, $\epsilon'$ and $\mu'$, that satisfies \eqref{transf}. Physically this can be understood as follows: Circulating currents can be described as magnetization, or alternatively as time-dependent polarization. The choice of parameters determines how much of the total induced current $\vek J$ is described by the magnetization vector and how much remains in the polarization. Clearly, only the transversal ($\perp$) part of the induced current can be associated with magnetization; the parallel ($\parallel$) part must remain in the polarization.

In the Landau--Lifshitz formulation, i.e. when the induced current is described solely by the permittivity, we can prove that for passive media \cite{landau_lifshitz_edcm}, 
\be\label{LLpassivity}
\im\epsilon'>0 \text{ for } \omega>0.
\ee
Using \eqref{epst} this means that \cite{pitaevskii12}
\be
\im\epsilon'_\perp = \im\epsilon_\perp - \frac{k^2c^2}{\omega^2}\im \mu^{-1}> 0.
\ee
Thus
\be\label{losscond}
 \im\mu > -\im\epsilon_\perp \frac{\omega^2|\mu|^2}{k^2c^2}.
\ee
In other words, the ambiguity for $\epsilon_\perp$ and $\mu$ makes it possible to define permeabilities or transverse permittivities with negative imaginary parts, even for passive media, as long as \eqref{losscond} is satisfied.

\section{Homogeneous conductors as diamagnetic media}\label{sec:homcond}
Inspired by the split-ring resonator metamaterial \cite{pendry99}, we realize that it is not always obvious which choice of parameters that is most ``physical''. One may take the view that a split-ring resonator metamaterial, made of nonmagnetic constituents, should be described by parameters $\epsilon'\neq 1$ and $\mu'=1$. This is a perfectly valid and natural choice \cite{silveirinha09}, given that there is no microscopic magnetization in the medium. However, metamaterial research has shown that it is convenient to describe the circulating currents using a macroscopic magnetization vector. This amounts to using a set of parameters $\epsilon$ and $\mu$, where $\mu\neq 1$.

As another well known example, we may consider a superconductor. Here there are two extremes \cite{roseinnes,orlando91}: Either the induced current is described explicitly, or the transverse current is described in the form of an effective magnetization. In the latter case, it is argued that one has diamagnetism. Note however, that the latter description may be somewhat dangerous, as the superconductor, being an intrinsically spatially nondispersive medium, here is described by a nonlocal permeability.
Also, the transverse current is absorbed into a magnetization vector, so e.g. in the second London equation the current must be expressed as $i\vek k\times\vek M$.

We now consider the superconductor example in more detail. If the induced current is described by the permittivity, the relative permeability is $\mu'=1$, and the superconductor can be modeled by a two-fluid model in which the conductivity has two terms. The permittivity is \cite{orlando91}
\be\label{epscond}
\epsilon' = 1-\frac{c^2}{\lambda^2\omega^2} - \frac{\omega_\text{p}^2}{\omega^2+i\omega\Gamma}
\ee
Here the second term describes the supercurrent, $\lambda$ being the London penetration depth. The third term describes the normal current due to the Drude model, with plasma frequency $\omega_\text{p}$ and a positive parameter $\Gamma$. For a (non-super) conductor, we can set $\lambda=\infty$. 

It is also common to refer to a superconductor as diamagnetic. In this alternative picture, circulating currents in the superconductor is described by a magnetization. Then $\epsilon_\perp=1$, and $\mu$ is obtained from
\be\label{mucond}
1-\mu^{-1} = -\frac{1}{\lambda^2k^2}-\frac{\omega_\text{p}^2}{c^2k^2}\cdot\frac{\omega^2}{\omega^2+i\omega\Gamma}
\ee
The equivalence with the first set of parameters $\epsilon'$ and $\mu'$ is seen by substitution into \eqref{transf}.

We observe from \eqref{mucond} that the permeability is nonlocal. The superconductor acts as a perfect diamagnet in the limit $k\to 0$; however for spatially varying fields the diamagnetism is not perfect due to the finite London penetration depth $\lambda$.

By inspection we find that $\mu^{-1}=\mu^{-1}(\omega,\vek k)$, as given by \eqref{mucond}, is analytic in the upper half-plane $\im\omega>0$, for any fixed $k$. Also note that the relative permeability from \eqref{mucond} is defined and has meaning even for high frequencies, as long as the original $\epsilon'$ has meaning. However, we observe that $\mu(\omega,\vek k)$ does not tend to unity as $\omega\to\infty$, but rather tends to a real number $\mu(\infty,\vek k)$ between 0 and 1:
\be\label{asymptmu}
\mu^{-1}(\infty,\vek k)=1+\frac{1}{\lambda^2k^2}+\frac{\omega_\text{p}^2}{c^2k^2}
\ee
Therefore, from the conventional proof of the Kramers--Kronig relations \cite{landau_lifshitz_edcm}, $\mu^{-1}$ satisfies a Kramers--Kronig relation of the type
\be\label{KK}
\re\mu^{-1}(\omega,\vek k) - \mu^{-1}(\infty,\vek k)  = \frac{2}{\pi}\text{P}\int_0^\infty \frac{\im\mu^{-1}(x,\vek k)x\diff x}{x^2-\omega^2},
\ee
where P denotes the Cauchy principal value. 

It is perhaps surprising that $\mu(\omega,\vek k)$ does not tend to unity for high frequencies. Here it is important to remember that in the presence of sources, $\omega$ and $\vek k$ are generally not connected. For eigenmodal propagation where $\omega$ and $\vek k$ are connected by the dispersion relation, the relative permeability will indeed tend to unity for high frequencies (see Sec. II). We also recall that despite $\mu\not\to 1$, the magnetization vector will tend to zero even for a fixed $\vek k$.

Setting $\omega=0$ in \eqref{KK} we see that for our example, diamagnetism is indeed compatible with causality and\\ $\im\mu(\omega,\vek k)>0$. The only requirement from \eqref{KK} is that
\be\label{critdiam}
\mu(0,\vek k) > \mu(\infty,\vek k).
\ee

To compare with previous literature, we now prove that the polarization--magnetization ambiguity means that for the same diamagnetic medium, other analytic $\mu$'s can be defined, that tend to 1 for high frequencies. These functions get negative imaginary parts for some frequencies, while not violating the passivity requirement \cite{pitaevskii12,martin67}. Our findings therefore do not contradict previous results \cite{Silveirinha11,pitaevskii12,landau_lifshitz_edcm,martin67}. To achieve $\mu\to 1$ for $\omega\to\infty$ and fixed $k$, we let division into $\epsilon_\perp$ and $\mu$ be frequency-dependent, so that the medium is described solely by a permittivity for high frequencies. From the parameter $\epsilon'_\perp$ (in the Landau--Lifshitz formulation where $\mu'=1$), we define a new set of parameters:
\begin{subequations}
\begin{align}
\epsilon_\perp &= 1 + (1-\alpha)(\epsilon'_\perp-1), \\
1-\mu^{-1} &= \alpha \frac{\omega^2}{k^2c^2}(\epsilon'_\perp-1).
\end{align}
\end{subequations}
The new set $\epsilon_\perp$ and $\mu$ is equivalent to the original set $\epsilon'_\perp$ and $\mu'=1$ according to \eqref{transf}. The parameter $\alpha$ is a weight factor describing the amount of transversal permittivity placed into $\mu$. It is natural to require $0\leq\alpha\leq 1$; however, in principle, $\alpha$ may be a completely arbitrary complex-valued function of $\omega$ and $\vek k$. Taking an ideal plasma $\epsilon'=1-\omega_\text{p}^2/\omega^2$ as a simple example, we obtain
\begin{subequations}
\begin{align}
\epsilon_\perp &= 1 - (1-\alpha)\frac{\omega_\text{p}^2}{\omega^2}, \\
1-\mu^{-1} &= -\alpha \frac{\omega_\text{p}^2}{k^2c^2}.
\end{align}
\end{subequations}
Now, provided $\alpha=\alpha(\omega)\to 0$ for $\omega\to\infty$, we will get the asymptote $\mu\to 1$. To describe diamagnetism at low frequencies, we require that $\alpha(0)=1$. We want $\mu^{-1}$ to be analytic, so $\alpha(\omega)$ needs to be analytic. This involves making it complex-valued. From the Kramers--Kronig relations, or in particular \eqref{KK} for $\omega=0$, we know that the resulting function $\mu$ must have a negative imaginary part somewhere in the spectrum. Clearly, since the new set $\epsilon_\perp$ and $\mu$ is equivalent to the passive original set $\epsilon'_\perp$ and $\mu'=1$, the negative imaginary part does not contradict passivity, and \eqref{losscond} is satisfied.

\section{Metamaterial examples}\label{sec:met}
We will now consider periodic metamaterials made from nonmagnetic constituents, i.e., dielectrics and conductors. If the metamaterial inclusions are described by a position dependent, microscopic, relative permittivity $\varepsilon(\vek r)$, it is known that $\varepsilon(\vek r)\to 1$ as $\omega\to\infty$. Therefore, as $\omega\to\infty$ the electromagnetic field will tend to the solution we would have if the metamaterial is replaced by vacuum. With a source $\vek J_\text{ext}=\bar{\vek J}_\text{ext}\exp(i\vek k\cdot\vek r)$, where $\bar{\vek J}_\text{ext}\perp\vek k$, we obtain from \eqref{maxwellsol} in this limit:
\begin{subequations}\label{maxwellsolvac}
\begin{align}
\vek B &=\frac{i\mu_0\vek k\times\vek J_{\text{ext}}}{k^2-\frac{\omega^2}{c^2}}, \label{BperpvacJ}\\
\vek E &=\frac{i\omega\mu_0\vek J_{\text{ext}}}{k^2-\frac{\omega^2}{c^2}}. \label{EperpJvac}
\end{align}
\end{subequations}
Thus, for sufficiently large $\omega$, we can express the microscopic, induced current $\vek j=-i\omega(\varepsilon(\vek r)-1)\epsilon_0\vek E$ as
\be\label{microj}
\vek j = \frac{\bar{\vek J}_{\text{ext}} \frac{\omega^2}{c^2}}{k^2-\frac{\omega^2}{c^2}} (\varepsilon(\vek r)-1)\exp(i\vek k\cdot\vek r).
\ee

As discussed in Sec. \ref{sec:elmagpar}, the magnetic permeability can be defined in several ways. A natural alternative in the so-called Casimir formulation \cite{landau_lifshitz_edcm,dirdal2017,vinogradov2002} is to define the magnetization from the magnetic moment density:
\be\label{Mint}
\vek M=\frac{\e{i\vek k\cdot\vek r}}{2V}\int_V \vek r\times\vek j\,\diff^3 r,
\ee
where we can e.g. choose the origin in the center of the unit cell. Compared to e.g. \cite{alu11} we have included an extra factor $\e{i\vek k\cdot\vek r}$ to be consistent with the definition of the macroscopic fields \eqref{averagefield1}. From \eqref{defMB} we now have a definition of a permeability, which we in principle can use for all frequencies.

\begin{figure}[t]
\begin{tikzpicture} [scale=5]
\path [draw=none,fill=gray, fill opacity = 1] (0,0) circle (0.42);
\path [draw=none,fill=white, fill opacity = 1] (0,0) circle (0.24);
\draw[-] (-0.5,-0.5)--(-0.5,0.5)--(0.5,0.5)--(0.5,-0.5)--(-0.5,-0.5);
\node [] at (0,0.32) {$\varepsilon(\omega)$};
\node [] at (-0.35,0.45) {$\varepsilon=1$};
\draw[<->] (-0.5,-0.55)--(0.5,-0.55);
\node [] at (0,-0.6) {$a$};
\draw[->] (0,0.001)--(0.24,0.025);
\node [right] at (0.05,0.07) {$b_1$};
\draw[->] (0,-0.001)--(0.42,-0.025);
\node [right] at (0.05,-0.07) {$b_2$};
\end{tikzpicture}
\caption{The metamaterial unit cell  consists of a cylinder of inner radius $b_1$ and outer radius $b_2$. The cylinder has relative permittivity $\varepsilon(\omega)$ and is surrounded by vacuum.} \label{fig:unitcell} 
\end{figure}
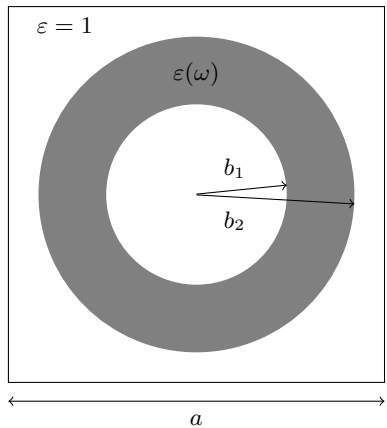

We first consider a 2d metamaterial consisting of quad\-ratic unit cells of area $a^2$. In the unit cell, there is a conducting ring of inner and outer radius $b_1$ and $b_2$, respectively, see Fig. \ref{fig:unitcell}. In the high-frequency range the relative permittivity of the conductor is approximated by a plasma response
\be
\varepsilon(\omega)=1-\frac{\omega_\text{p}^2}{\omega^2}.
\ee
By calculating the integral \eqref{Mint} under the assumption $ka\ll 1$, and using \eqref{defMB}, we find that in the high-frequency regime
\be\label{mulim}
1-\mu^{-1}(\infty) = -\frac{\pi\omega_\text{p}^2}{8c^2}\frac{b_2^4-b_1^4}{a^2}.
\ee
When the ring is seen as a cylinder in 3d, we must interpret $1-\mu^{-1}$ as the $(z,z)$ element of corresponding tensor. Eq. \eqref{mulim} shows that the relative permeability tends to a value between 0 and 1 as $\omega\to\infty$ while $ka$ is fixed ($ka\ll 1$).

As proved in Appendix \ref{appKK}, the function $\mu^{-1}(\omega,\vek k)$ resulting from \eqref{defMB} with \eqref{Mint} is analytic in the upper half-plane of complex frequency ($\im\omega>0$), for fixed $\vek k$. Thus $\mu^{-1}$ satisfies a Kramers--Kronig relation of the form \eqref{KK}.

\begin{figure}[t]
\begin{tikzpicture} [scale=5]
\node [right] at (0.6,1.1) {$y$};
\node [right] at (1.0,0.5) {$x$};
\draw [fill=white] (0,0)--(0,1)--(0.4,1)--(0.4,0)--(0,0);
\draw [fill=gray] (0.4,0)--(0.4,1)--(0.8,1)--(0.8,0)--(0.4,0);
\draw [<->] (0,0.050) -- (0.4,0.05);
\node [above] at (0.2,0.05) {$a/2$};
\node [] at (0.2,0.6) {$\varepsilon=1$};
\draw [<->] (0,-0.050) -- (0.8,-0.05);
\node [below] at (0.4,-0.05) {$a$};
\draw [->] (0.6,0) -- (0.6,1.1);
\draw [->]  (-0.1,0.5) -- (1.0,0.5);
\node [white] at (0.6,0.6) {$\varepsilon(\omega)$};
\node [above, white] at (0.6,0.05) {$a/2$};
\draw [<->, white] (0.4,0.050) -- (0.8,0.05);
\end{tikzpicture}
\caption{Unit cell of a layered medium (1d metamaterial).}\label{fig:1Dstruct}
\end{figure}
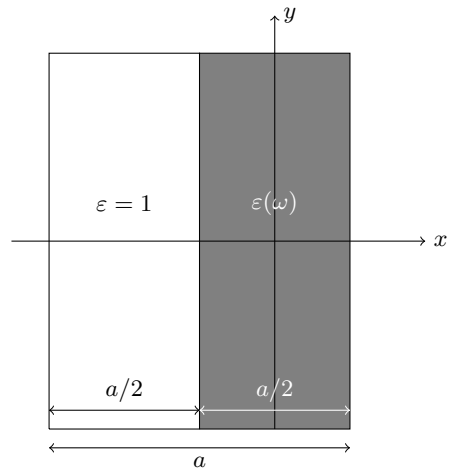

\begin{figure}[h]
\includegraphics[width=0.45\textwidth]{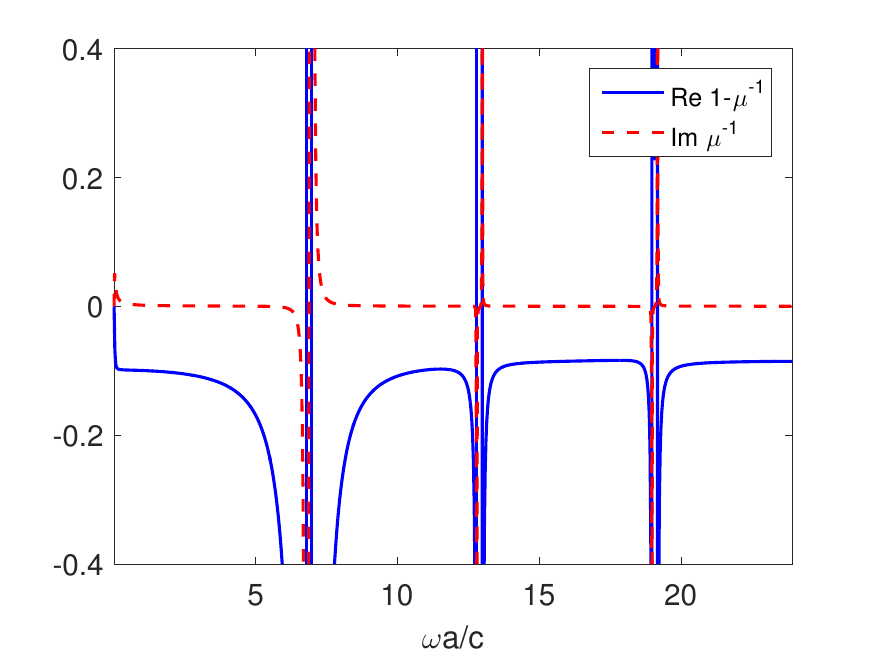}
\caption{$1-\mu^{-1}$ vs. frequency for $ka=0.1$. Several resonances are observed which become narrower with increasing frequency. At these resonances, $\im\mu$ changes sign, and one may think that the parameter is unphysical. \emph{However, although $\mu$ does not retain its usual interpretation as a local parameter, it is still physical}, in the sense that the definition based on averaged magnetic moment of the unit cell (\eqref{Mint} and \eqref{defMB}) applies. Given this definition of $\mu$, it is not surprising that the imaginary part can be negative; it only means a phase shift of the local field. The Landau--Lifshitz permittivity (which includes all electric and magnetic multipoles) is verified to have positive imaginary part, as required for a passive medium. For large frequencies, above the plasma frequency $\omega_\text{p} a/c=4.0$, we note that $1-\mu^{-1}$ approaches the asymptote given by \eqref{eq:Asym1Dmu}: $1-\mu^{-1}=-0.086$ is observed for $\omega a/c=24$. Notice also that $1-\mu^{-1}$ approaches zero as $\omega \to 0$. }\label{fig:DispMu1D} 
\end{figure}

We next consider a 1d metamaterial consisting of alternating layers of vacuum and copper (Fig. \ref{fig:1Dstruct}). The microscopic permittivity of copper is described by a Drude model:
\begin{align}
    \varepsilon(\omega) = 1 - \frac{\omega_\text{p}^2}{\omega^2 + i \omega \Gamma},
\end{align} 
where $\omega_\text{p}=1.20\cdot 10^{16}~$s$^{-1}$ and $\Gamma=5.24\cdot 10^{13}~$s$^{-1}$ \cite{Ordal:83}. Making use of boundary conditions and the Bloch property of the fields, the microscopic fields can be found by use of transfer matrices, thereby allowing for straightforward calculation of the parameter $1-\mu^{-1}$. Fig. \ref{fig:DispMu1D} displays a plot of $1-\mu^{-1}$ vs. frequency for $a=100\,$ nm and $ka=0.1$. Repeated resonances are observed, which become narrower with increasing frequency. Except the resonances, which will be discussed in more detail below, $1-\mu^{-1}$ is observed to approach an asymptote unequal to zero. Similarly to \eqref{mulim}, the asymptote for this 1d structure can be calculated to be
\begin{align}\label{eq:Asym1Dmu}
1-\mu^{-1}(\infty) &= -\frac{1}{192}\frac{\omega_\text{p}^2}{c^2}a^2 \\
&= -0.084. \nonumber
\end{align} 
This asymptote corresponds well with the dispersion of $1-\mu^{-1}$ shown in Fig. \ref{fig:DispMu1D}.

There are several resonances, resulting from the periodic structure. At the resonances, we observe that $\im\mu$ changes sign. Clearly, we cannot view $\epsilon$ and $\mu$ as local parameters describing the metamaterial's response in this range. Nevertheless, the parameter $\mu$ has a certain physical meaning, being defined from the averaged magnetic moment of the unit cell (\eqref{Mint} and \eqref{defMB}). A negative imaginary part is a result of a phase shift of the local field in the cell. The fact that the parameter is well defined and physical for all frequencies, makes it possible to use the usual Kramers--Kronig relations without truncation or any other modifications (Appendix \ref{appKK}). We have also computed the permittivity $\epsilon'$ in the Landau--Lifshitz formulation (not shown here). It satisfies $\im\epsilon'>0$ for all frequencies, as required by passivity \eqref{LLpassivity}.

\section{Conclusion}
It is well known that diamagnetism is an effect related to spatial dispersion, although the medium can behave spatially nondispersive for restricted wavenumber spectra. It turns out that diamagnetism at zero frequency is compatible with Kramers--Kronig relations and a positive $\im\mu$ for all frequencies, as the asymptote of the relative permeability for high frequencies and fixed $\vek k$ can be different from 1. We point out that such an asymptote is permitted by Maxwell's equations and causality, and provide analytical and numerical examples of associated diamagnetic media and metamaterials.

\appendix
\section{Analyticity and Kramers--Kronig relations}\label{appKK}
It is interesting to explore the analytic properties of the electromagnetic parameters \cite{landau_lifshitz_edcm,agranovich84,dolgov81,alu11c}. If we use the La\-ndau--Lifshitz formulation in which the medium is desc\-ribed solely by a permittivity $\epsilon'$ (and $\mu'=1$), it is usually assumed that $\epsilon'$ is an analytic function of $\omega$ \cite{landau_lifshitz_edcm,agranovich84}. This follows by regarding the electric field as the excitation and the displacement field as the response. However, as pointed out in \cite{dolgov81}, such an argument is not compelling since the electric field includes the response of the medium. 

In \eqref{maxwellsol} the fields are expressed from the sources, which means that it is straightforward to identify the response functions. Treating the source $\vek J_\text{ext}$ as the excitation, and the electric field as the response, it follows from \eqref{EparJ} that $1/\epsilon'_\parallel(\omega,\vek k)$ is analytic in the upper half-plane $\im\omega>0$ for fixed $\vek k$. Moreover, from \eqref{EperpJ}
\be\label{resp}
R(\omega)\equiv \frac{i\omega\mu_0}{k^2-\frac{\omega^2}{c^2}\epsilon'_\perp} 
\ee
must be analytic in the upper half-plane. Even though $R(\omega)$ is analytic, it is not entirely obvious that $\epsilon'_\perp$ is. Since $R(\omega)$ is analytic there, any zero of $R(\omega)$ is of finite order. We can write 
\be
\epsilon'_\perp(\omega,\vek k)=\frac{R(\omega)k^2-i\omega\mu_0}{R(\omega)\omega^2/c^2},
\ee
and thus $\epsilon'_\perp(\omega,\vek k)$ is analytic except possibly of poles. In \cite{alu11c} it is proved that $\epsilon'(\omega,\vek k)$ does not contain any poles in the upper half-plane, for a metamaterial made of causal constituents with analytic permittivities in the upper half-plane. With the additional property $\epsilon'\to 1$ as $\omega\to\infty$, the Kramers--Kronig relations are established.

We now leave the Landau--Lifshitz formulation and describe the medium with both $\epsilon$ and $\mu$. Due to the presence of both functions $\epsilon_\perp(\omega,\vek k)$ and $\mu(\omega,\vek k)$ in \eqref{BperpJ}-\eqref{EperpJ}, they are not necessarily analytic functions separately. For example, we may choose to describe the transversal current by the permeability up to a given frequency, and abruptly describe it using the permittivity for higher frequencies. Although a somewhat artificial choice, it demonstrates that an extra condition is required to establish analyticity for $\epsilon_\perp$ and $\mu$. 

For the permeability resulting from the magnetization \eqref{Mint}, we can prove analyticity for $\mu^{-1}$ as follows. If the metamaterial is described with a single permittivity tensor $\epsilon'=\epsilon'(\omega,\vek k)$ (Landau--Lifshitz formulation), the magnetic flux density is given by
\be
\vek B =\frac{i\mu_0\vek k\times\vek J_{\text{ext}}}{k^2-\frac{\omega^2}{c^2}\epsilon'_\perp}, \label{BperpJLL}\\
\ee
analogously to \eqref{BperpJ}. Substituting \eqref{Mint} and \eqref{BperpJLL} into \eqref{defMB}, we have
\be
\frac{\mu_0}{2V}\int_V \vek r\times\vek j\,\diff^3 r = (1-\vek\mu^{-1})\frac{i\mu_0\vek k\times\bar{\vek J}_{\text{ext}}}{k^2-\frac{\omega^2}{c^2}\epsilon'_\perp}.
\ee
Choosing a source with $\vek k=k\vekh x$ and $\bar{\vek J}_\text{ext}=\bar{J}_\text{ext}\vekh y$, we find for the $(z,z)$-element of $(1-\vek\mu^{-1})$:
\be
1-\mu^{-1} = \frac{1}{\bar{J}_{\text{ext}}}\frac{k^2-\frac{\omega^2}{c^2}\epsilon'_\perp}{2ikV} \vekh z\cdot\int_V \vek r\times\vek j\,\diff^3 r.
\ee
The source $\bar{J}_\text{ext}$ can be chosen to be analytic in the upper half-plane. It can also be chosen zero-free. The function $\epsilon'_\perp$ is analytic in the upper half-plane, provided the metamaterial is made of causal constituents \cite{alu11c}. Clearly the microscopic, induced current $\vek j$ is analytic in the upper half-plane, since it is causally related to the source. It therefore follows that $\mu^{-1}$ is analytic in the upper half-plane.

With the analyticity and the asymptotic behavior, Kra\-mers--Kronig relations \eqref{KK} for $\mu^{-1}$ can finally be stated. 

\def\cprime{$'$}

\end{document}